\documentclass[aps,prl,twocolumn,psfig,showpacs]{revtex4-1}
\usepackage{amsfonts}
\usepackage{amsmath}
\usepackage[dvips]{graphicx}
\usepackage{bm}
\usepackage{amssymb}
\usepackage{times}
\usepackage{dcolumn}
\usepackage{cases}
\usepackage{txfonts}
\usepackage{color}

\RequirePackage[hyperindex,colorlinks,bookmarksnumbered,plainpages=true]{hyperref}
\hypersetup{colorlinks,linkcolor=blue,urlcolor=blue,citecolor=blue}
\usepackage{hyperref}

\newcommand{\tj}[6]{ \begin{pmatrix}
  #1 & #2 & #3 \\
  #4 & #5 & #6
\end{pmatrix}}

\usepackage{ulem}%only for the command \sout = scrap
\renewcommand{\emph}[1]{\textit{#1}}%needed, if package  "ulem" is used

\definecolor{darkgreen}{rgb}{0,0.5,0}
\definecolor{darkblue}{rgb}{0,0,0.5}
\definecolor{darkred}{rgb}{.7,0,0}
\definecolor{purple}{rgb}{0.35,0,0.35}
\definecolor{orange}{rgb}{1,0.5,0}
\definecolor{grey}{rgb}{.6,.6,.6}

\newcommand{\Fig}[1]{Fig.~\ref{#1}}

\begin{document}
\title{Simplex valence-bond crystal in the spin-1 kagome Heisenberg antiferromagnet}
\author{Tao Liu$^1$}
\author{Wei Li$^{2,3}$}
\email{w.li@physik.lmu.de}
\author{Andreas Weichselbaum$^2$}
\author{Jan von Delft$^2$}
\author{Gang Su$^1$}
\email{gsu@ucas.ac.cn}
\affiliation{$^{1}$Theoretical Condensed Matter Physics and Computational Materials Physics Laboratory, School of Physics, University of Chinese Academy of Sciences, Beijing 100049, China
\linebreak $^{2}$Physics Department, Arnold Sommerfeld Center for Theoretical Physics, and Center for NanoScience, Ludwig-Maximilians-Universit\"at, 80333 Munich, Germany
\linebreak $^{3}$Department of Physics, Beihang University, Beijing 100191, China}
\date{\today}

\begin{abstract}
We investigate the ground state properties of a spin-1 kagome antiferromagnetic Heisenberg model using tensor-network (TN) methods. We obtain the energy per site {$e_0=-1.41090(2)$ with $D^*=8$ multiplets retained (i.e., a bond dimension of $D=24$), and $e_0=-1.4116(4)$ from large-$D$ extrapolation,} by accurate TN calculations directly in the thermodynamic limit. The symmetry between the two kinds of triangles is spontaneously broken, with a relative energy difference of $\delta \approx$ 19\%, i.e, there is a trimerization (simplex) valence-bond order in the ground state. The spin-spin, dimer-dimer, and chirality-chirality correlation functions are found to decay exponentially with a rather short correlation length, showing that the ground state is gapped. We thus identify the ground state be a simplex valence-bond crystal (SVBC). We also discuss the spin-1 bilinear-biquadratic Heisenberg model on a kagome lattice, and determine its ground state phase diagram. Moreover, we implement non-abelian symmetries, here spin SU(2), in the TN algorithm, which improves the efficiency greatly and provides insight into the tensor structures.
\end{abstract}

\pacs{75.10.Jm, 75.10.Kt, 05.10.Cc}
\maketitle

\textit{Introduction.---} Geometrical frustration, as a particularly interesting phenomenon in quantum antiferromagnets, has raised enormous interest recently \cite{Misguich}. It arises when any classical (Ising)  spin configuration cannot satisfy simultaneously all the local terms in the Hamiltonian, which leads to a macroscopic degeneracy and thus greatly enhances quantum fluctuations. Frustration might melt semiclassical spin orders (including magnetic or valence bond order, etc.), driving the system into an exotic quantum state called quantum spin liquid \cite{Lee, Balents}. Some typical frustrated antiferromagnets include the spin-1/2 and spin-1 Heisenberg models on the triangular lattice \cite{White-2007, Lauchli}, spin-1/2 $J_1$-$J_2$ square \cite{Capriotti-2001, Hu-2013, Jiang-2012, Gong-2013a, Wang}, and the pyrochlore \cite{Bramwell-2001} lattices. Among others, the spin-1/2 kagome Heisenberg antiferromagnetic (KHAF) model is one of the most intriguing frustrated models: its ground state is widely believed to be a spin liquid \cite{White-2011, Schollwoeck-2012, Jiang-2012b, Gong-2013, Poilblanc-2012, Carrasquilla, Xie-2014}, but its nature is still under debate \cite{Iqbal}.

KHAF models with higher spins \cite{Goetze} are less well-studied, despite their physical realizations in experiments, e.g., m-MPYNN$\cdot$BF4 \cite{Awaga, Wada, Watanabe, Matsushita, Lawes, Hamaguchi}  {and YCa$_3$(VO)$_3$(BO$_3$)$_4$ \cite{Miller}}, where the measurements reveal a gapped nonmagnetic state with only short-range spin ordering. Interesting variational wavefunctions have been proposed for the relevant spin-1 KHAF model, for instance, the static or resonating Affleck-Kennedy-Lieb-Tasaki (AKLT) loop state states \cite{HYao-2010, Li-2014, ZCai-2007}, and the hexagon-singlet solid state \cite{Hida_HSS}, etc, yielding some preliminary advances towards understanding the nature of the ground state. Notably, Cai \textit{et al.} considered a fully trimerized variational wavefunction on the kagome lattice \cite{ZCai-2007}, with all the spin-1's in each $A$ (or $B$) triangle forming a singlet (trimerization). However, its corresponding variational energy for the spin-1 KHAF model is $e_0=-1$ per site, much higher than that of the topologically ordered resonating AKLT-loop state (a quantum equal-weight superposition of all possible AKLT-loop coverings, $e_0\approx -1.27$) \cite{Li-2014}. The nature of the ground state of the spin-1 KHAF model is still an open question.

\begin{figure}
\includegraphics[width=0.9\linewidth,clip]{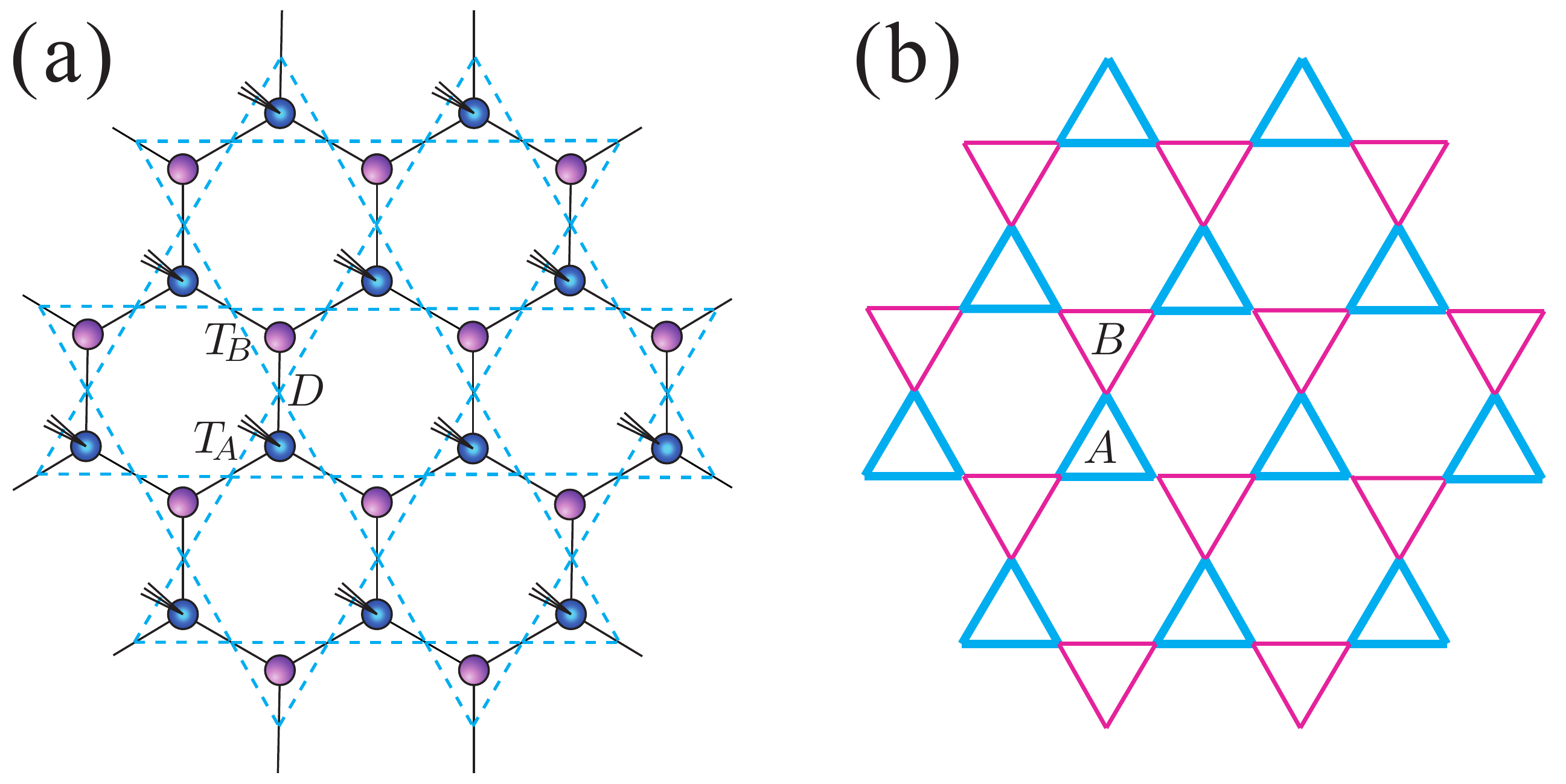}
\caption{(Color online) (a) Kagome lattice (dotted lines) and the initial setup of the tensor-network wavefunction (solid lines). $D$ is the bond dimension, and $T_A$ ($T_B$) are triangle tensors, {with which the physical indices can be associated for convenience.} (b) Illustration of the simplex valence-bond crystal. The two kinds of triangles or ``simplexes" [of type $A$ (blue) and $B$ (pink)] have different energies, and a lattice inversion symmetry is spontaneously broken.}
\label{fig1}
\end{figure}

In this work, we employ state-of-the-art tensor network (TN) algorithms \cite{Jiang-2008, Li-2012, Liu-2014} based on the projected entangled-pair state (PEPS) to study the properties of spin-1 KHAF model, and determine the variational ground state energy as $e_0\simeq-1.41$ on an infinitely large two-dimensional (2D) lattice [\Fig{fig1}(a)]. Lattice inversion (reflection) symmetry is found to be broken, where the two kinds of triangles (or simplexes) have different energies [\Fig{fig1}(b)]. We thus call the ground state a simplex valence-bond crystal (SVBC). We also consider the spin-1 bilinear-biquadratic (BLBQ) Heisenberg model, and obtain its ground state phase diagram, where we find an extended SVBC phase and observe a quantum phase transition between the SVBC and ferro-quadrupolar phases at $\theta_c \simeq -0.04 \pi$. Some of our results were obtained with an SU(2)-invariant implementation of PEPS, coded using the QSpace tensor library \cite{Weich}, which greatly reduces the costs (supplementary materials).

\textit{Model and Method.---} We consider the quantum spin-1 KHAF model with only nearest-neighbor isotropic exchange interactions (i.e., Hamiltonian (\ref{4}) with $\theta=0$). We use the PEPS as a wavefunction ansatz \cite{Verstraete-2004}, and invoke an imaginary-time evolution (through the Trotter-Suzuki decomposition \cite{fn-trotter}) for optimizations. The initial hexagonal TN [\Fig{fig1}(a)] consists of tensors $T_A$ and $T_B$, associated with all $A$ and $B$ triangles of the lattice, respectively. {Such a TN ansatz has also been employed to study the spin-1/2 KHAF model \cite{Xie-2014}.} 

\begin{figure}
\includegraphics[width=0.92\linewidth,clip]{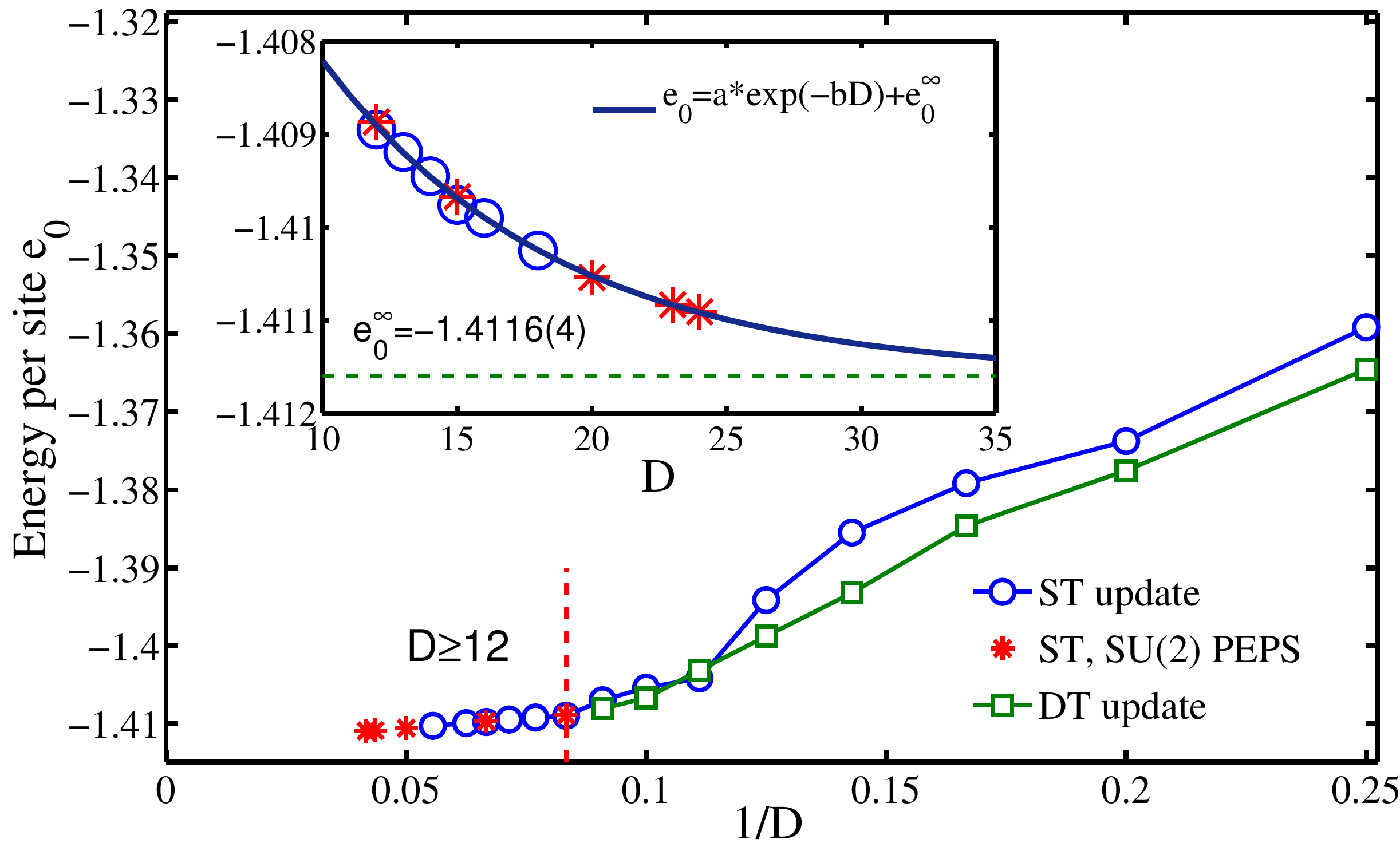}
\caption{(Color online) {The variational ground state energy per site $e_0$ is shown versus $1/D$, obtained from iPEPS contractions [with and without implementing SU(2) symmetry] on the infinite kagome lattice, using both ST and DT update schemes. The inset shows that the $D\geq12$ data (i.e., left-hand-side of the dashed line) converge exponentially to the infinite $D$ limit, which is extrapolated as $e_0^{\infty}=-1.4116(4)$. The convergence of $e_0$ versus truncation parameters $d_c$ have always been checked (supplementary material), and the data above are obtained with $d_c=40\sim60$ and $100\sim120$ for plain and SU(2) iPEPS contractions, respectively.}}
\label{fig2}
\end{figure}

After each step of the imaginary-time evolution, we have to reallocate the three physical indices (from $T_A$ to $T_B$, or the other way round) and truncate the bond state space. Here we use the single-triangle (ST) or double-triangle (DT) update schemes for truncations (supplementary materials), following Refs. \onlinecite{ODTNS, Liu-2014, Xie-2014, Li-2012}. {We find good agreement between ST and DT calculations once the bond dimension $D$ is sufficiently large (see, e.g., Figs.~\ref{fig2} and \ref{fig-trimer}), indicating that ST calculations are sufficient to accurately capture the ground-state properties.} 

\begin{figure}
\includegraphics[width=1\linewidth,clip]{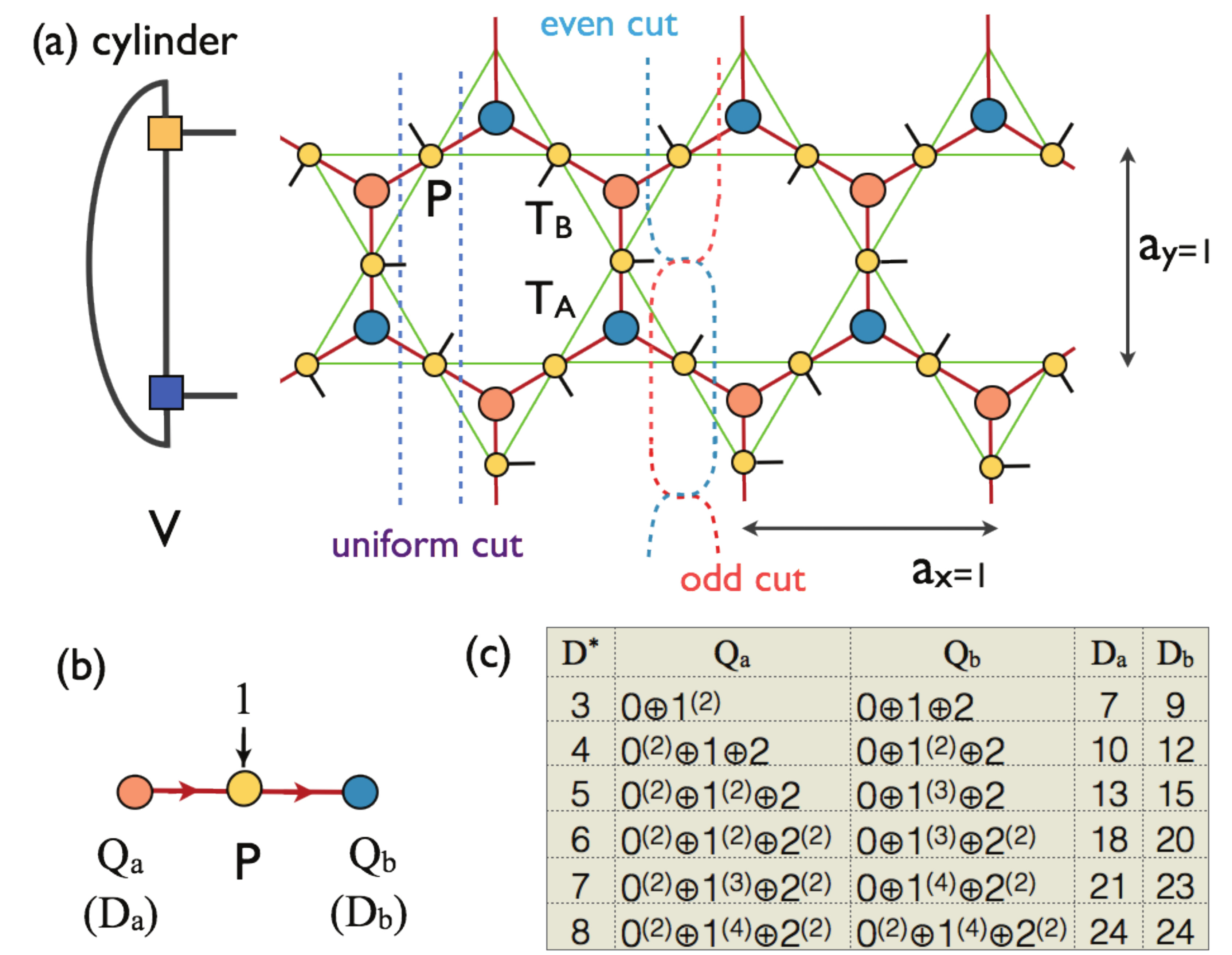}
\caption{(Color online) (a) Illustration of the cylinders. For XC(YC) geometries, X(Y) direction is with periodic boundary condition, and length unit $a_{x}$($a_y$), $L$ is the circumference. (b) Implementation of SU(2) symmetry in local tensors, the arrows indicate how the spin multiplets are fused together \cite{fn-symm}. The table in (c) shows the specific spin representations $Q_{a,b}$ (and corresponding plain bond dimensions $D_{a,b}$) of the optimized tensors for various $D^*$ (i.e., number of kept bond multiplets). {Here $S^{(m)}$ means $m$ multiplets with spin $S$.}}
\label{fig3}
\end{figure}

{We has also implemented SU(2) symmetry in the TN algorithms, greatly improving the numerical efficiency. To this end, we employed the tensor library \mbox{QSpace} \cite{Weich}, which implements non-abelian symmetries in TNs in an efficient and transparent framework. We have run data for $D^*=3\sim8$, where $D^*$ is the number of multiplets retained on the geometric bonds [see \Fig{fig3}(b,c)], as compared to the actual number of states $D$. In the imaginary-time evolution, we only specify the number $D^*$ of retained multiplets, while the representations with respect to SU(2) spin symmetry are free to change during the optimization process, and eventually converge to the integer ones specified in \Fig{fig3}(c).}

Given the optimized tensors {(with or without SU(2) symmetry)}, we consider two geometries for evaluating the expectation values: (a) an infinitely large 2D lattice and (b) an infinitely long cylinder with finite circumference (\Fig{fig3}). For case (a), we adopt the infinite PEPS (iPEPS) technique \cite{Jordan-2008, Orus-2009, Orus-2008} to contract the double-layer TN, with boundary matrix product state (MPS) retaining $d_c$ bond states. For case (b), we wrap the TNs on the X- or Y-cylinders (denoted XC or YC in previous work on kagome cylinder \cite{White-2011, Schollwoeck-2012}), and contract the boundary vector [$V$ in \Fig{fig3}(a)] with a column of tensors, repeating this process until convergence is reached. 

\textit{Ground state energy and valence-bond crystal.--- } \Fig{fig2} presents our results of energy per site $e_0$. The inset shows that $e_0$'s are well converged with retaining $d_c\geq40$ bond states in the boundary MPS. The main panel, where $d_c=40$, shows that the energy decreases monotonically with increasing bond dimension $D$, {reaching $e_0 = -1.41090(2)$ for $D^*=8$ (i.e., $D=24$). In the inset of Fig.~\ref{fig2}, we find that the $D\geq12$ data are well in the exponential convergence region, and the corresponding fit suggests $e_0^{\infty} =-1.4116(4)$ in the infinite $D$ limit.} This constitutes our best estimate of the ground state energy in the thermodynamic limit.

\begin{figure}
\includegraphics[width=0.85\linewidth,clip]{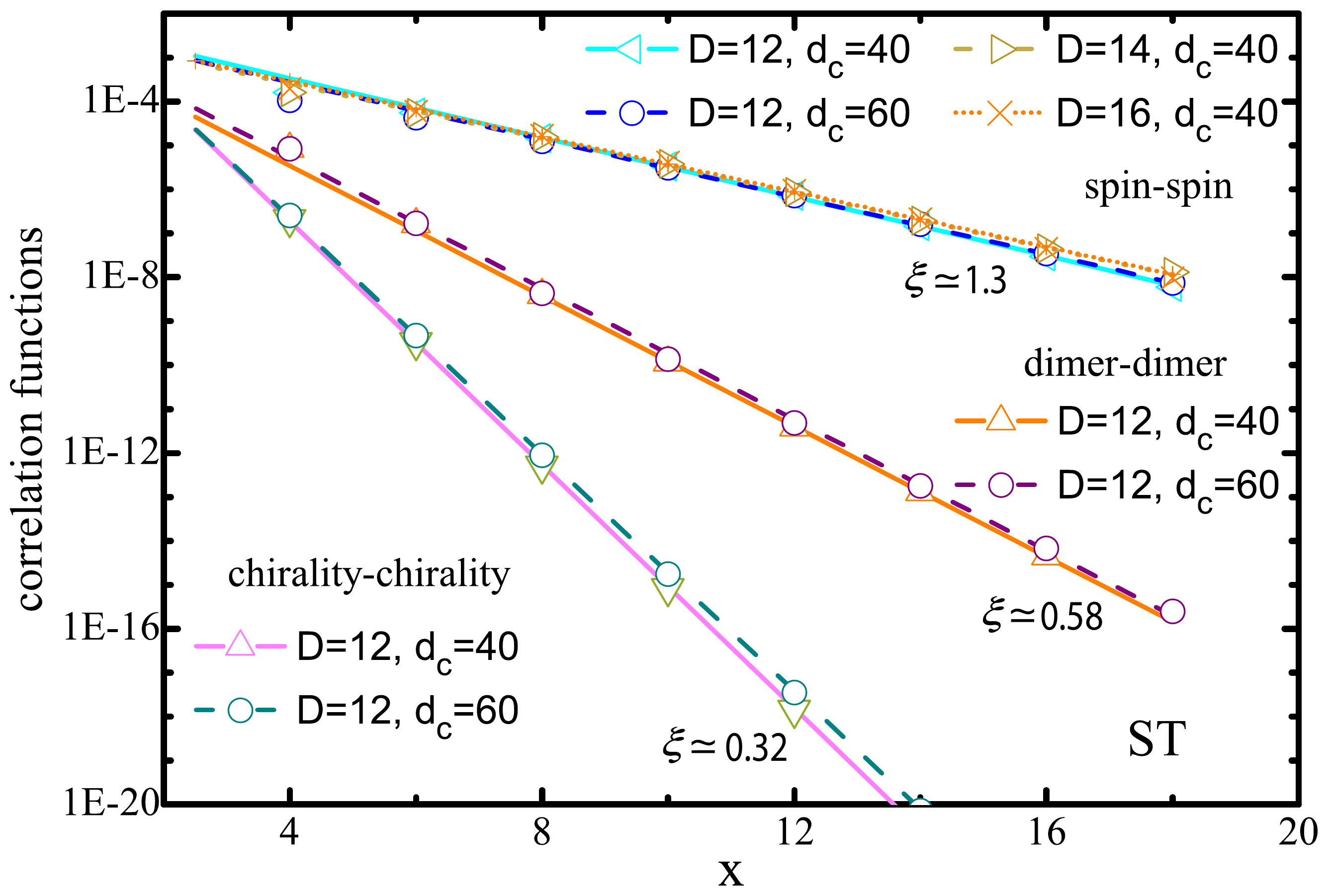}
\caption{(Color online) Spatial dependence of various correlation functions (symbols) on a log-linear scale, together with exponential fits $y=c \exp{(-x/\xi)}$, with $\xi$ indicated with each line. The correlation functions are calculated by iPEPS. $x$ is the distance between triangles with length unit $a_x$ (see \Fig{fig3}(a)). Note that the square of the converged $\langle S^z_i S^z_{i+1}\rangle \neq 0$ has been subtracted in the definition of dimer-dimer correlations.}
\label{fig-cf}
\end{figure}

\begin{figure}[htpb]
\includegraphics[width=1\columnwidth]{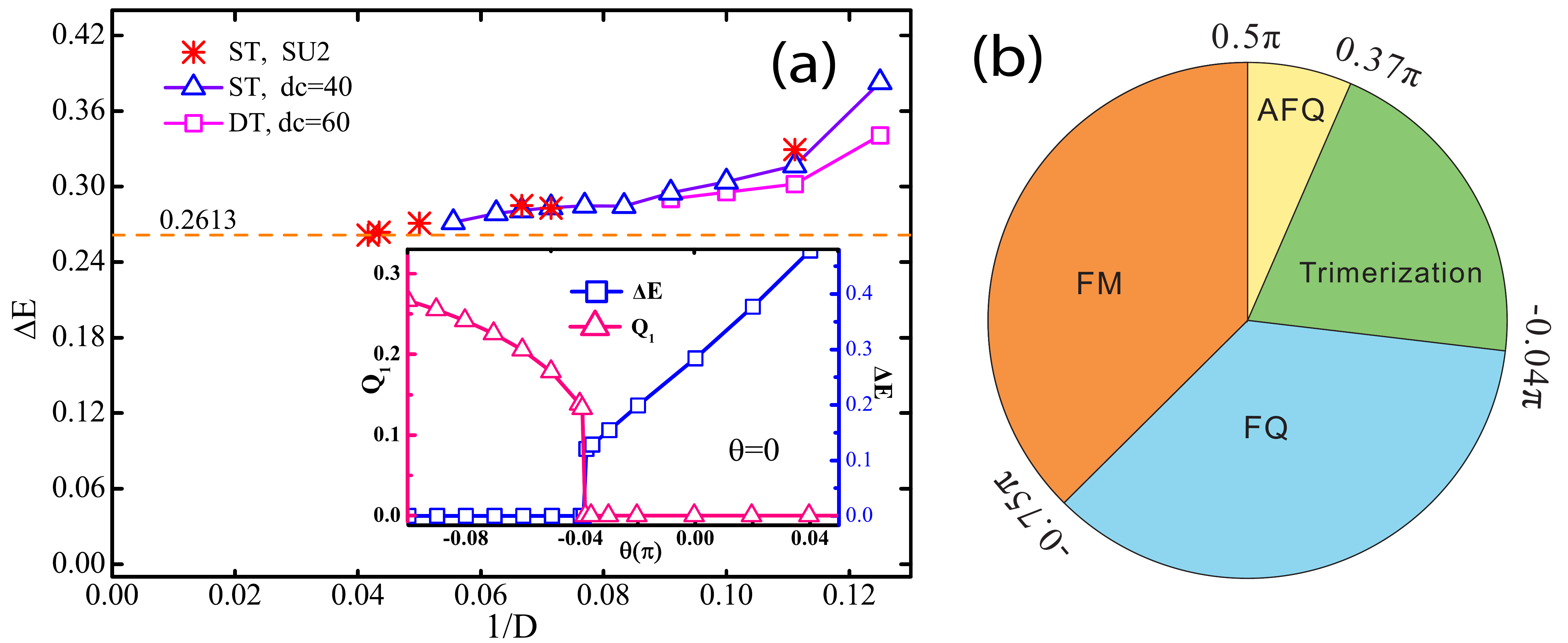}
\caption{(Color online) (a) Energy difference {between $A$ and $B$ triangles,} $\Delta E = 2(E_A-E_B)/3$, where $E_{A(B)} = 9\langle S_i^z S_{i+1}^z\rangle_{A(B)}$, evaluated at the Heisenberg point ($\theta=0$) with the iPEPS contraction, and plotted versus $1/D$, {which show clearly a non-vanishing value ($\delta = \Delta E/e_0 \approx 19\%$ for $D=24$)}. The minimal bond dimension needed to capture the SVBC order is $D\gtrsim7$ [or $D^*=3$, see table in \Fig{fig3}(c)]. For smaller $D$, $\Delta E$ vanishes, and hence is not shown here. The inset shows that $\Delta E$ vanishes when $\theta < -0.04 \pi$, where ferromagnetic quadrupolar order ($Q_1$) sets in. (b) Ground state phase diagram of the spin-1 BLBQ model on the kagome lattice. }
\label{fig-trimer}
\end{figure}

In \Fig{fig-cf}, we show the spin-spin, dimer-dimer, and chiral correlation functions, all evaluated between equivalent sites of two triangles of the same kind, say, $A$ triangles. The spin-spin correlation function is defined by $\langle S^z_i S_j^z\rangle$, and the dimer-dimer one by $\langle D_i D_j \rangle = \langle(S_{i}^{z} S_{i+1}^{z})\cdot (S_{j}^{z} S_{j+1}^{z})\rangle-\langle S_{i}^z S_{i+1}^{z}\rangle \cdot \langle S_{j}^{z}  S_{j+1}^{z}\rangle$, where $i$ and $j$ belong to different triangles. The chiral correlation function is defined as $\langle C_{m} \, C_{n}\rangle=\langle [\bold{S}_{m_{1}}\cdot (\bold{S}_{m_{2}}\times \bold{S}_{m_{3}})] [{\bold{S}}_{n_{1}} \cdot ({\bold{S}}_{n_{2}}\times {\bold{S}}_{n_{3}})]\rangle$, where $m,n$ label positions of two triangles, and $m_i,n_i$ label the positions of the three sites within a triangle. \Fig{fig-cf} shows that all these correlation functions decay exponentially, implying that the ground state of spin-1 KHAF model is non-magnetic and gapped.

\Fig{fig-trimer} shows the energy difference $\Delta E =\frac{2}{3} |E_A-E_B| $ between $A$ and $B$ triangles, as a function of $D$. The fact of non-vanishing $\Delta E$ means that the ground state spontaneously breaks lattice inversion symmetry. Note, although our method is initially biased in its treatment of $A$ and $B$ triangles in the ST update, by the end of the projections we reduce the Trotter slice to $10^{-5}$, restoring the equivalence between the two kinds of triangles. Besides th e ST update, we have also employed the DT update, where the two triangles are treated on equal footing, for determining the ground state. The quantitative agreement between the ST and DT results in \Fig{fig-trimer} confirm the stability of the spontaneous trimerization order.

\begin{figure}
\includegraphics[width=1\linewidth,clip]{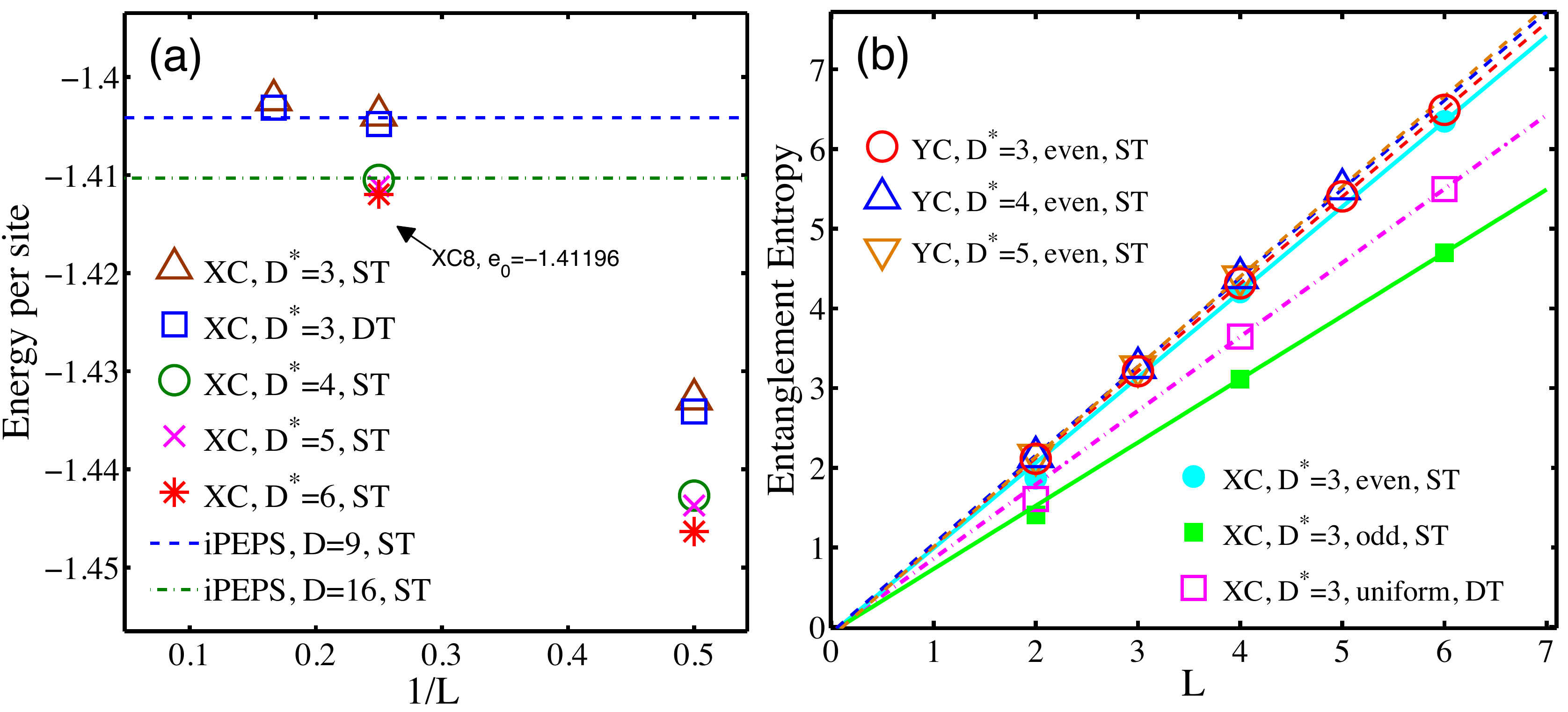}
\caption{(Color online) (a) Energy per site and (b) von Neumann entanglement entropies of the tensor-network variational wavefunctions on cylinders. {The X(Y)C geometry is shown in Fig.~\ref{fig3}(a), and $L=2, 4, 6$ means infinite X(Y)C4, 8, 12 cylinders, respectively.}}
\label{fig-cyl}
\end{figure}

\textit{Bilinear-biquadratic Heisenberg model.---} We also studied the spin-1 BLBQ Heisenberg model with Hamiltonian
\begin{equation}\label{4}
  H=\sum\limits_{<ij>}[\cos\theta \,  (\bold{S}_{i}\cdot \bold{S}_{j})+ \sin\theta \, (\bold{S}_{i} \cdot \bold{S}_{j})^{2}],
\end{equation}
which recovers the KHAF model when $\theta=0$. When we tune $\theta$ away from the Heisenberg point, we see that the SVBC state belongs to an extended phase. The results are shown in the inset of \Fig{fig-trimer}(a). The energy differences are verified to be robust for various $\theta$'s. Interestingly, when we tune $\theta$ to the negative side, a phase transition occurs at the transition point $\theta_c\simeq-0.04$, where the trimerization vanishes,  and the system turns into a ferro-quadrupolar (FQ) phase, with $Q_1 = \langle  S_x^2 -S_y^2  \rangle \neq 0$.

In \Fig{fig-trimer}(b) shows the ground state phase diagram of the spin-1 kagome BLBQ Heisenberg model obtained by exploring other $\theta$ values. There are four phases in total: a FQ phase ($-3/4 \pi < \theta < -0.04 \pi$), a SVBC phase ($-0.04 \pi < \theta < 0.37 \pi$), an antiferro-quadrupolar (AFQ) phase ($0.37\pi < \theta < 1/2 \pi$,  $\bold{Q}_{tot} = \sum_{i \in \triangle} \bold{Q_i} = \bold{0}$, but $\bold{Q_i} \neq \bold{0}$), and a ferromagetic (FM) phase ($1/2 \pi < \theta < 5/4 \pi$). Note that the SU(3) point ($\theta=\pi/4$) lies in the SVBC phase, thus the SU(3) Heisenberg model also has a trimerized ground state. This observation is in agreement with a previous study of the SU(3) model \cite{Corboz, Arovas}. Note also that Fig. \ref{fig-trimer}(b) is similar to the phase diagram of the spin-1 BLBQ model on a triangular lattice \cite{Lauchli-2006}, but the antiferromagnetic phase there replaced by the SVBC phase, and the SU(3) point there is no longer a phase transition point here.

\textit{Exact contractions with SU(2) PEPS.---} The implementation of non-abelian symmetries leads to a huge numerical gain in efficiency, especially in the contractions of double-layer TNs. For example, we are able to perform exact contractions on a cylinder as large as XC12 for the $D^*=3$ state, thanks to a factor of 340 reduction in the memory (from about 2000 GB to 6 GB, see supplementary materials). A very promising future application would be in iPEPS full update which scales as $D^{10\sim12}$ \cite{Jordan-2008}; due to the very large exponent, the numerical gain from tracking $D^*$ multiplets rather than $D$ individual states per bond can be expected to be huge.
 
\Fig{fig-cyl}(a) shows the energy expectation values up to XC12 ($L=6$). For the $D^*=3$ case, the DT offers slightly better energy compared to the ST data. Thanks to the implementation of SU(2) symmetry, we are able to evaluate an optimized {$D^*=6$ state on XC8 ($L=4$), yielding a variational energy of $e_0 = -1.41196$, a variational upper bound of $e_0$ on a given cylinder, and it agrees well with the iPEPS results in Fig.~\ref{fig2}.} In addition, trimerization can also be clearly identified in the optimized $D^*=3,4,5,6$ states, again with a relative difference $\sim$20\%. 

\textit{Entanglement entropy.---} We cut the cylinder PEPS into two halves, and evaluate the von Neumann entropy \cite{Cirac-2011}, $S=-\rm{Tr} [\rho \log(\rho)]$, fitting it to $S \simeq c L - \gamma$. For the topological states, $\gamma$ extrapolates to a nonzero constant \cite{Jiang-2012b}, called the topological entanglement entropy (TEE) \cite{Kitaev-2006, Levin-2006}. \Fig{fig-cyl}(b) shows the von Neumann entropies of {$D^*=3$ states (obtained with ST or DT update) on XC and YC geometries with $L=2,4,6$.} In the ST update case, owing to the PEPS construction, the cylinder can be cut in two inequivalent ways, called even or odd cut [see \Fig{fig3}(a)]. In the DT case, where the unit cell tensor is larger, we can cut the cylinder in a uniform way [\Fig{fig3}(a)]. {Besides the $D^*=3$ state, Fig. \ref{fig3}(b) also shows the entanglement entropies of $D^*=4,5$ states evaluated on various YC geometries; the ``even" cut there means the entropies are calculated when the physical indices are associated with $T_A$ in Fig. \ref{fig3}(a).} All the entanglement results extrapolate to $\gamma \simeq 0$, suggesting a topologically trivial state. 

\textit{Conclusions and discussion.---} We find the ground state of the spin-1 KHAF model to be a gapped SVBC, evidenced by the spontaneous lattice symmetry breaking between two neighboring triangles. More generally, the striking contrast between the ground states of the spin-1/2 and spin-1 kagome antiferromagnets, spin liquid vs. trimerized crystal, raises an interesting question: does spin parity matter for higher-spin  kagome antiferromagnets, too? 
An important technical innovation of our work is the explicit implementation of SU(2) symmetry in our PEPS-based algorithms; this not only enhances their numerical performance, but also provides us useful information about the bond multiplets. To be concrete, the SVBC state and the fully trimerized (trivial) state share some common virtual-spin representations and fusion channels in the tensors. This suggests that the two states are adiabatically connected. In the supplementary materials we show numerically that this is indeed the case.

Lastly, we address some remarks on the relation to experimental observations. The susceptibility measurements of the organic spin-1 magnet m-MPYNN$\cdot$BF4 reveal a gapped, nonmagnetic ground state \cite{Wada, Watanabe, Matsushita}, consistent with our SVBC picture, which is nonmagnetic and has a spin gap. However, the specific heat measurement shows a round peak at $T/2J'$ $\sim$ 1/2 ($2J'\approx 3K$, the coupling strength), suggesting only a short-range ordering. This observation suggests that other complications in the materials (like next-nearest couplings, distortions, single-ion anisotropy, etc) should be taken into account, which we leave for a further study.

\textit{Acknowledgement.---} WL was indebted to Hong-Hao Tu, Meng Cheng, Shuo Yang, Zi Cai, and Tomotoshi Nishino for stimulating discussions. TL thanks Guang-Zhao Qin for his help in polishing the schematic plot. We acknowledge Hong-Chen Jiang and Shou-Shu Gong for discussions about the DMRG calculations of the same model. This work was supported in part by the MOST of China (Grant  No. 2012CB932900 and No. 2013CB933401), and the Strategic Priority Research Program of the Chinese Academy of Sciences (Grant No. XDB07010100). WL was also supported by the DFG through SFB-TR12 and NIM, and acknowledges the hospitality of the Max-Planck Institute for Quantum Optics, where part of the work has been performed. AW further acknowledges support by DFG grant WE-4819/1-1.

\textit{Note added.---} {We have noticed three recent preprints,} two on density matrix renormalization group study of the same model \cite{Lauchli-2014, Nishimoto-2014}, and the other on tensor-network study of magnetization curves of spin-1 kagome model and others \cite{Poilblanc-2014}: {two with conclusions consistent with ours \cite{Lauchli-2014,Poilblanc-2014}, while the other proposed a different ground state \cite{Nishimoto-2014}.}

\onecolumngrid
\vspace{2cm}
\begin{center}
{\bf\large Supplementary Material}
\end{center}
\vspace{0.2cm}
\setcounter{equation}{0}
\renewcommand{\theequation}{S\arabic{equation}}  
\setcounter{figure}{0} 
\renewcommand{\thefigure}{S\arabic{figure}}

\onecolumngrid

\section{The single- and double-triangle updates}
In this part we describe the method employed to optimize the tensors, the single- and double-triangle update, and its interpretation as a Bethe-lattice approximation. 

\begin{figure}
\includegraphics[width=0.95\linewidth,clip]{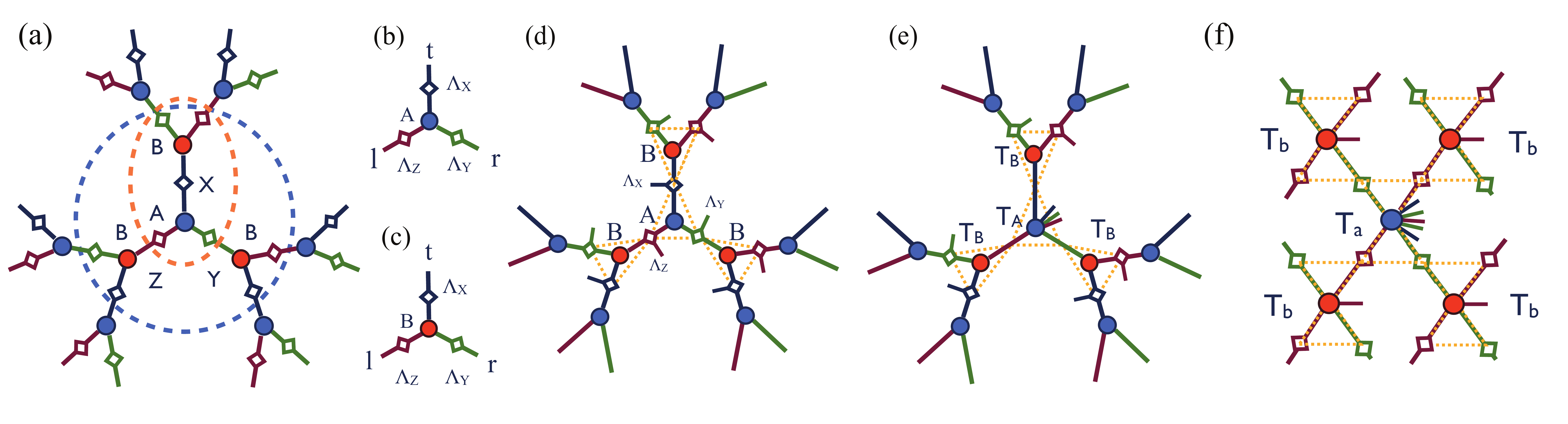}
\caption{(Color online) (a) The Bethe lattice with coordination number three.  (b,c) Local tensors determined by single-triangle update. (d)  The Husimi lattice with triangle motifs, the corresponding tensor network consists of triangle tensors $A, B$ and tensors $\Lambda_{x,y,z}$ (with physical indices) located on the geometric bonds. (e) Associate three physical indices with one triangle tensor $T_A$, we get the tensors used in practical calculations. (f) A larger cluster is adopted in the double-triangle update scheme, where the tensor $T_a$($T_b$) with four virtual indices represents two triangles instead of one.}
\label{figS1}
\end{figure}

We start from the so-called simple update on the honeycomb lattice, whose corresponding Beth lattice is shown in Fig. \ref{figS1}(a). We treat the couplings between the four sites within the dashed blue circle exactly, and update the subsystem (during the imaginary-time evolution) with the help of environments. To be specific, the entanglement between the cluster and the rest of the lattice is well approximated by the objects $\Lambda_{x,y,z}$ on the geometric bonds, which play an important role in the bond truncations. The $\Lambda$'s are determined iteratively and self-consistently during the imaginary time evolution. 

In one dimension (1D), this method, dubbed simple update \cite{Jiang-2008}, recovers the iTEBD method \cite{Orus-2008}, is a quasi-optimal way to truncate the geometric bonds and update the tensors. In higher spatial dimensions, the simple update generates optimal wavefuncions only on the Bethe lattice with an infinite Hausdorff dimension, thus it can also be called the Bethe-lattice approximation \cite{Li-2012}. The simple update is quite efficient, however for intermediate dimensions, say 2D, unfortunately, it does not guarantee to always produce optimal tensor-network variational wavefunction. In case simple update fails to generate accurate results, one has to resort to the so-called full update \cite{Jordan-2008, Orus-2009}, which is much more costly both in time and memory. Therefore, there is always a trade off between using expensive full update with only small bond dimension $D$ available, and using simple update with much larger $D$. The method of choice is usually model dependent: in case the ground state of the model possesses rather local properties, say the SVBC ground state of the spin-1 KHAF model studied in the present paper, simple update turns out to be a very good optimization scheme since it allows for large bond dimension $D$, which is necessary for producing accurate results.

For the kagome lattice, the corresponding ``tree" structure lattice consists of corner-sharing triangles [see dashed triangles in Fig. \ref{figS1} (d,e)]. This lattice, dual to the Bethe lattice, is called the Husimi lattice. Associating a tensor with each triangle, we again obtain a Bethe-lattice tensor network representation for the model defined on the Husimi lattice, and thus can perform simple update very conveniently \cite{Liu-2014}. Notice that now the $\Lambda$'s have additional physical indices, and the triangle projection operator contains three sites. In practice, we do not store $\Lambda$'s explicitly, but rather associate them with $A$ ($B$) and obtain the triangle tensors $T_A$ ($T_B$) [Fig. \ref{figS1}(e)].

During the imaginary-time evolutions, after absorbing a three-site projection operator $e^{-\tau h_{A}}$ ($h_\alpha$ the triangle Hamiltonian, $\alpha \in \{A,B\}$) with the tensor $T_A$, we make a decomposition
\begin{equation}
\label{3}
  (T_A)^{p_{1},p_{2},p_{3}}_{\tilde{x},\tilde{y},\tilde{z}} = \sum_{x,y,z}(\tilde{T}_A)_{x,y,z}(\Lambda_{x})^{p_{1}}_{x,\tilde{x}}(\Lambda_{y})^{p_{2}}_{y,\tilde{y}}(\Lambda_{z})^{p_{3}}_{z,\tilde{z}}, 
\end{equation}
where $\Lambda_{x,y,z}$ can be obtained by higher-order singular value decomposition \cite{Xie-2014}.
Then, $T_A$ is replaced by a tensor $\tilde{T}_A$ without physical indices. Subsequentially, we associate the three $\Lambda$ tensors with $T_B$, and update it with $\tilde{T}_B$ using
\begin{equation}\label{4}
  (\tilde{T}_B)^{p_{1},p_{2},p_{3}}_{\tilde{x},\tilde{y},\tilde{z}} = \sum_{x,y,z}(T_B)_{x,y,z}(\Lambda_{x})^{p_{1}}_{x,\tilde{x}}(\Lambda_{y})^{p_{2}}_{y,\tilde{y}}(\Lambda_{z})^{p_{3}}_{z,\tilde{z}}.
\end{equation}
After this tensor-network transformation, the three physical indices have moved from $T_A$ to $\tilde{T}_B$ and we can proceed to act with the projection $e^{-\tau h_{B}}$ on $\tilde{T}_B$. This procedure can be repeated until the tensors $T_A$ and $T_B$ converge. 

{Besides the above single-triangle (ST) simple update scheme, we have also used a double-triangle (DT) update scheme, with an enlarged unit cell tensor [see Fig. \ref{figS1}(f), called 5-PESS in Ref. \onlinecite{Xie-2014}], to get better optimization of the tensors. In practice, we perform projections until the energy expectations converge within a prescribed accuracy of, say, $10^{-10}$.}

Having obtained the local tensors, we rewire the tensors on an infinite 2D lattice or on a cylinder, and evaluate the observables using accurate iPEPS/cylinder PEPS techniques. It turns out that, as long as the bond dimension $D$ is large enough, the DT calculations are in good agreement with the ST data, suggesting that for the spin-1 KHAF model ST update is sufficient to accurately capture the ground-state physics.

{\section{An adiabatic connection to a fully trimerized state}}
{In this section, we discuss the connection between the SVBC ground state of spin-1 KHAF and the fully trimerized state. Their connection can be realized in a heuristic way, by looking at the fusion channels of the SU(2)-invariant tensors. The fully trimerized state has a very simple SU(2) representation. The triangle tensor is proportional to a Wigner 3j symbol:}
\begin{equation}
T_{ m_1, m_2, m_3 }^{ j_{1}, j_2, j_3} = \tj{j_1}{j_2}{j_3}{m_1}{m_2}{m_3},
\end{equation}
{indicating how to properly fuse three multiplets $\{j_1,j_2,j_3\}$ into a rotationally invariant singlet. For the fully trimerized state, $T_A = T^{1,1,1}$, and $T_B = T^{0,0,0}$, because if the three spin-1's on the corners of every $A$-triangle are fully bound into a trimer singlet (described by the tensor $T_A^{1,1,1}$), there is no spin content left on the corners of every $B$-triangle (described by the tensor $T_B^{0,0,0}$). In addition, the projection tensor $P$ connecting $T_A$ and $T_B$ is as}

\begin{equation}
    P^{j_1,j_2} =
   \begin{cases}
   1 &\mbox{if } j_1+j_2=1 \\
   0 &\mbox{otherwise},
   \end{cases}
\end{equation}
{because every site connecting $A$- and $B$-triangles hosts a total spin-1.}

{On the other hand, if one looks carefully into the optimized tensor ($D^*=3,4,5$), one can identify, among many other fusion channels, the $T^{1,1,1}$ channel on one triangle, say $T_A$, and $T^{0,0,0}$ on the other, $T_B$. Take $D^*=3$ state for instance, it turns out that a total of four channels contribute to $T_A$, namely $T^{1,1,1}$, $T^{1,0,1}$, $T^{1,1,0}$, and $T^{0,1,1}$, but no $T^{0,0,0}$; while 11 channels contribute to $T_B$ tensor, including $T^{0,0,0}$, $T^{1,1,0}$, $T^{1,1,2}$, $T^{2,2,0}$, $T^{2,2,2}$, etc, but no $T^{1,1,1}$. This observation, true also for higher $D^*$, suggest that by gradually reducing to zero the weight of all the fusion channels in the tensors, except for $T^{1,1,1}$ on $T_A$ and $T^{0,0,0}$ on $T_B$, one would cross over from the SVBC state to the fully trimerized state.}

{In order to confirm this conjecture, and numerically verify that these two state belong to the same phase, we consider the distorted Heisenberg model \cite{Hida_HSS}, with different coupling constants in two different kinds of triangles as}

\begin{equation}
H_{st} = J_A h_{\triangle, A} + J_B h_{\triangle, B}, 
\end{equation}
{where $h_{\triangle, A(B)}$ are spin-spin Heisenberg Hamiltonians (for three sites) in a triangle $A$($B$), $J_A$ and $J_B$ are the coupling constants.}

\begin{figure}
\includegraphics[width=0.8\linewidth,clip]{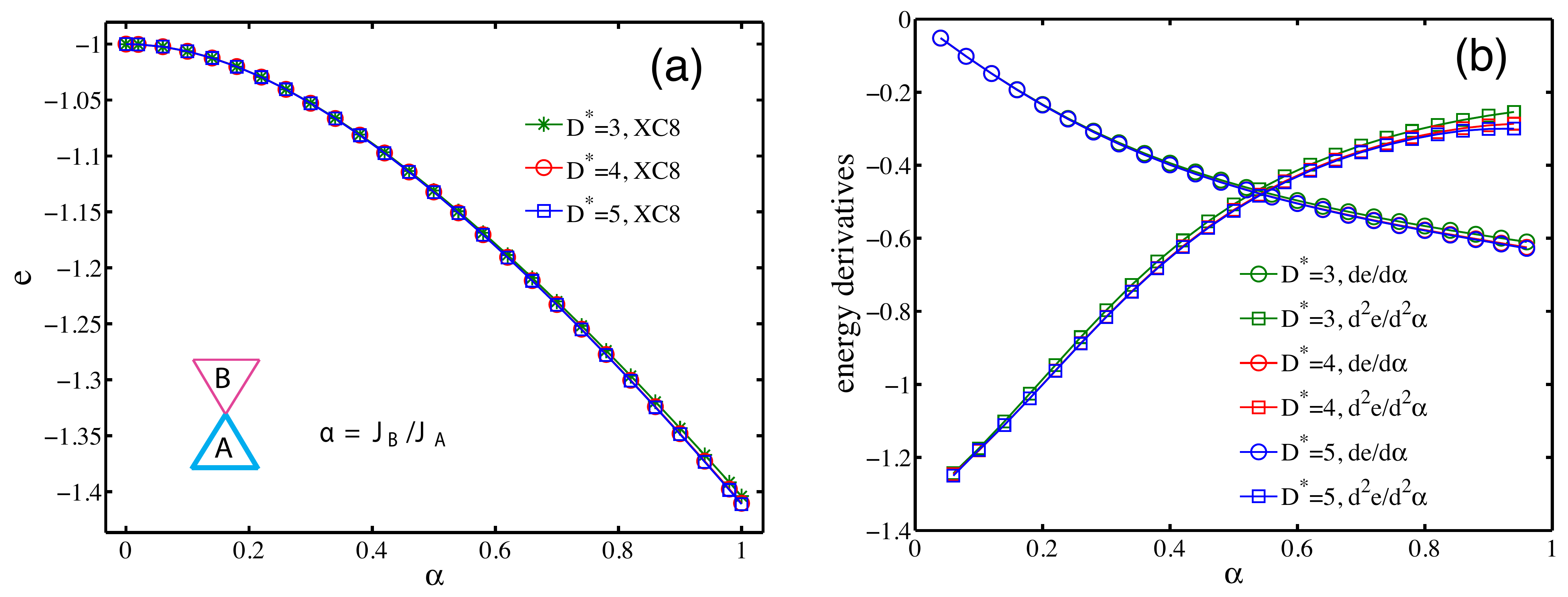}
\caption{(Color online) {(a) The energy per site $e$ and (b) its first- and second-order derivatives, respect to parameter $\alpha = J_B/J_A$, the ratio between the spin couplings in triangles $A$ and $B$.}}
\label{figS2}
\end{figure}

{Tuning parameter $\alpha = J_B/J_A$ from 0 to 1, we have fully trimerized ground state at $\alpha=0$, and recover at $\alpha=1$ the homogeneous Heisenberg model. We employ SU(2) PEPS calculations, the results obtained by retaining bond multiplets $D^*=3,4,5$ are shown in Fig. \ref{figS2}. The energy per site $e$ and its derivatives with respect to $\alpha$ (up to second order) are continuous and smooth, which leads to the conclusion that the simplex valence bond crystal (SVBC) ground state of the spin-1 kagome is in the same phase as the fully trimerized state.}

$\\$
\section{Non-abelian symmetries in the tensor networks}
{We have implemented non-abelian symmetries in our tensor network algorithms, using the QSpace tensor library \cite{Weich}, which is a generic, efficient, and transparent implementation of non-abelian symmetries in the tensors. The QSpace can be applied to matrix product state (MPS), tree tensor network (TTN), projected entangled-pair state (PEPS), or any other kind of tensor networks. To keep the notation compact, we take the MPS as an example, to elaborate the basic idea of QSpace. The basis transformation of MPS can be ``factorized" into two parts, involving tensor $A$ containing reduced matrix elements, and tensors $C$ containing Clebsch-Gordan coefficients (CGCs)
\begin{equation}
| \tilde{Q} \tilde{n}; \tilde{Q}_z \rangle = \sum_{Q n, Q_z} \sum_{q l,q_z} (A_{Q,\tilde{Q}}^q)_{n, \tilde{n}}^{l}(C_{Q, \tilde{Q}}^q)_{Q_z, \tilde{Q}_z}^{q_z} |Q n; Q_z\rangle |q l; q_z\rangle.
\label{eq:QSpace-State}
\end{equation}}
{Here $Qn$, $\tilde{Q}\tilde{n}$ are the composite multiplet indices of the ancillas, and $ql$ stands for the physical space. $Q = [Q_1, Q_2, ..., Q_m]$ is the multiplet label (for $m$ different symmetries $\mathcal{S} \equiv \bigotimes_{\lambda=1}^m \mathcal{S}^{\lambda}$; e.g., for SU(2) $\otimes$ U(1),  $m=2$), $n$($\tilde{n}$, $l$) distinguishes different multiplets with the same $Q$($\tilde{Q}$, $q$), and $Q_z$($\tilde{Q_z}$, $q_z$) distinguish the individual states within a multiplet $Q$($\tilde{Q}$, $q$), respectively. The tensor $(C_{Q, \tilde{Q}}^q)_{Q_z, \tilde{Q}_z}^{q_z} = \bigotimes_{i=1}^{m} (\tilde{Q}_i \tilde{Q}_{i z} | Q_i Q_{i z}; q_i q_{i z})$ stores the CGCs.}

{The reduced tensors $A$ take care of the fusion on the level of multiplets, e.g. input multiplets $(Q n)$ and a local space $(q l)$ into output multiplets $(\tilde{Q} \tilde{n})$ (usually truncated) in the reduced multiplet level; the CGC tensors $C$ take care of the underlying mathematical symmetry structure. The QSpace is a very useful concept not only for describing states, but also for expressing irreducible tensor operators. According to the Wigner-Eckart (WE) theorem, their matrix elements can be expressed as
\begin{equation}
\langle \tilde{Q} \tilde{Q}_z |  F^q_{q_z} | Q, Q_z \rangle = (F_{\tilde{Q}, Q}^q)^{[1]}_{\tilde{n},n} \cdot (C_{Q, \tilde{Q}}^q)_{Q_z, \tilde{Q}_z}^{q_z},
\end{equation}}
{which shares the same structure as the basis transformation in Eq. \ref{eq:QSpace-State}, i.e., also a product of $ (F_{\tilde{Q}, Q}^q)^{[1]}_{\tilde{n}, n}$ in the reduced multiplet space and $C_{Q, \tilde{Q}}^q$ in the CGC space (note the inverse order of $Q, \tilde{Q}$ in the CGCs, owing to the WE theorem).}

{With the QSpace tensor library, it is straightforward  to implement non-abelian symmetries such as SU(N) or Sp(2n) generally, for example, the SU(2)$_{\textrm{spin}}$ symmetry for Heisenberg spin chains, and the SU(2)$_{\textrm{spin}}$ $\otimes$ SU(2)$_{\textrm{charge}}$ symmetry for fermonic chains, etc. To develop a QSpace SU(2) DMRG code, one only needs to find the reduced $A$-tensors variationally as in plain DMRG, while the underlying CGC space ($C$-tensors) are fully determined by symmetry. Because the relevant $A$-tensors only work on the reduced multiplet space (whose dimension is much smaller than that of the full tensor), it leads to a huge gain in numerical efficiency.}

{In spatial dimensions larger than one, the QSpace framework can be employed too, in principle, for any kind of tensor network (in general, then $A$- and $C$-tensors have more than three indices). However, when implementing non-abelian symmetries, here SU(2)$_{\mbox{spin}}$, in two-dimensional (2D) PEPS, one faces a conceptual problem, due to the occurrence of loops in the tensor network. An analogous problem arises in MPS-base algorithms, when switching from open to periodic boundary conditions (PBC): For open boundary conditions, SU(2) symmetry can be implemented in the so-called ``renormalization" picture, where one adds one site after another to the system, and the symmetry labels ($Q n$ of virtual particles) are generated naturally for the orthonormal bases associated with each bond. However, when PBC are adopted, i.e., the MPS forms a closed ring, a tricky point arises as one loses the orthonormality of the bond bases (i.e., it is no longer possible to bring the MPS into a canonical form), which seemingly messes up the concept of symmetry labels. The situation becomes even ``worse" in 2D PEPS, where there are many closed loops in the tensor network. That is to say, the ``renormalization" picture for introducing symmetry labels in 2D PEPS needs a more careful consideration.}

{On the other hand, if one looks at some SU(2)-invariant states which have a simple 2D PEPS representation, like spin-1/2 resonating valence-bond state \cite{Poilblanc-2012}, or spin-1 resonating Affleck-Kennedy-Lieb-Tasaki state \cite{Li-2014}, their local tensors are all SU(2)-invariant. This can be understood in terms of the so-called ``projection" picture: Virtual particles are introduced around each physical site, and every virtual particle forms a singlet-pair with a nearest-neighbor virtual particle on the same bond; one then introduces an SU(2)-invariant projection operator, mapping the ancillas associated with a given site into the desired physical degree of freedom of that site. Following this ``projection" picture, it is possible to implement non-abelian symmetry in PEPS, by generalizing the simple RVB or RAL state to a general state whose total wavefunction is a symmetry eigenstate. Furthermore, one can argue that this ``projection" picture is complete in principle, i.e., it can cover (generate) any symmetry eigenstates. A more detailed discussion will appear elsewhere.}

\begin{figure}
\includegraphics[width=0.65\linewidth,clip]{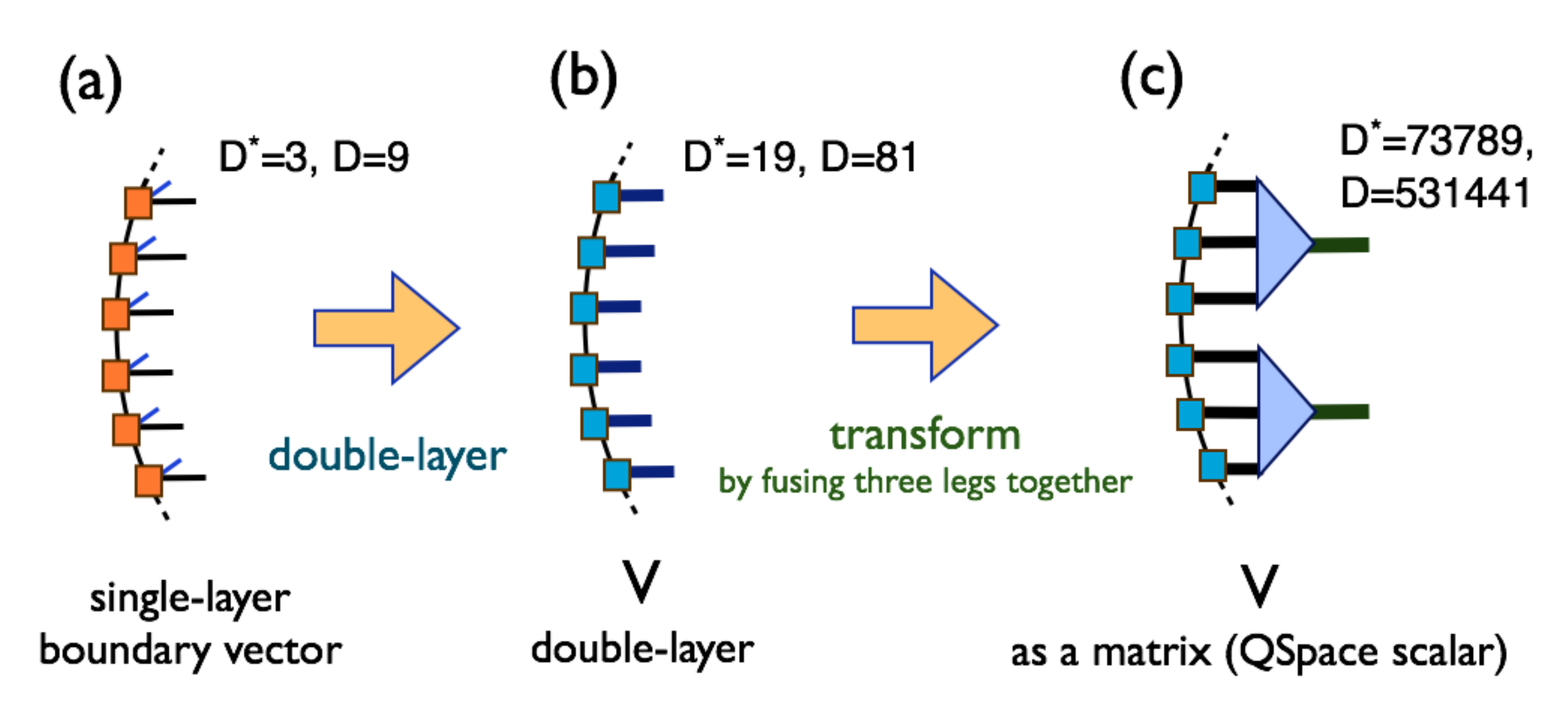}
\caption{(Color online) {(a) Single-layer boundary vector of  XC12 cylinder, with $D^*=3$ multiplets (corresponding to $D=9$) on every geometric bond. (b) After taking the inner product, we obtain the boundary vector $V$ in the double-layer tensor network, which has $D^*=19$ ($D=81$) fat geometric indices per fat geometric bond. (c) To estimate the memory costs, we can transform $V$ into a QSpace scalar (matrix), by fusing every three legs into a composite one.}}
\label{figS3}
\end{figure}

{Once SU(2) symmetry has been implemented in the PEPS algorithm, the numerical benefit is huge and thus very promising for future applications. To obtain the results shown in the main text, we have implemented SU(2) symmetry in both the imaginary-time evolution and exact/iPEPS contraction codes. This allows us to perform exact contractions of cylinders as large as XC12 for the $D^*=3$ state, whereas an exact contraction was not possible for plain PEPS of the corresponding $D=9$ state. In this specific case, the SU(2)-invariant boundary vector $V$ (see Fig. \ref{figS3}), which can be stored as an SU(2) scalar object in QSpace, by fusing three double-layer legs together [see Fig. \ref{figS3}(c)], corresponds to a $73789 \times 73789$ matrix (reduced multiplet dimension $D^*$). Its plain counterpart $V_p$ is as large as $531 441 \times 531 441$ (plain dimension $D$), meaning that there is roughly a factor of $ 7.2^2 \approx 51$ memory cost reduction: from 2104.3 GB (full matrix) to around 40.5 GB (QSpace scalar). Furthermore, $V$ has a block-diagonal structure in the reduced multiplet space, which further reduces the memory from the naive estimate of 40.5 GB down to 6.21 GB in practice, i.e., by a factor of about 6.5. It is this total factor of about 340 (338.8) reduction in memory cost which enables us to evaluate interesting observables, such as the energy per site, of the $D^*=3$ ($D=9$) state on XC12. Furthermore, when one evaluates more complex quantities, such as the entanglement entropy, the numerical benefits are even greater, because the computational cost reduction then may scale as $7.2^3 \approx 373$ or so (one has to decompose $V$ or perform matrix multiplications, etc). }

{Another demonstration of both the feasibility and the benefits of implementing non-abelian symmetries is the iPEPS contraction. Its computational cost scales as O($D^{10\sim12}$) on a square lattice model \cite{Jordan-2008}, which is a power law but has a large exponent, which thus prevents from retaining relatively large $D$ in the PEPS simulations. When non-abelian symmetries have been implemented, we can track $D^*$ multiplets per bond instead of $D$ individual states, and greatly enhance the bond dimensions [in principle it doubles (or triples) the bond dimensions, in the case of SU(2)]. The implementation of non-abelian symmetries in tensor networks is just in its beginning, and we do think it would show its full power for using tensor networks to tackle those ``hard" problems in condensed matter physics, like the frustrated antiferromagnets and interacting fermion models.}

\begin{figure}
\includegraphics[width=0.5\linewidth,clip]{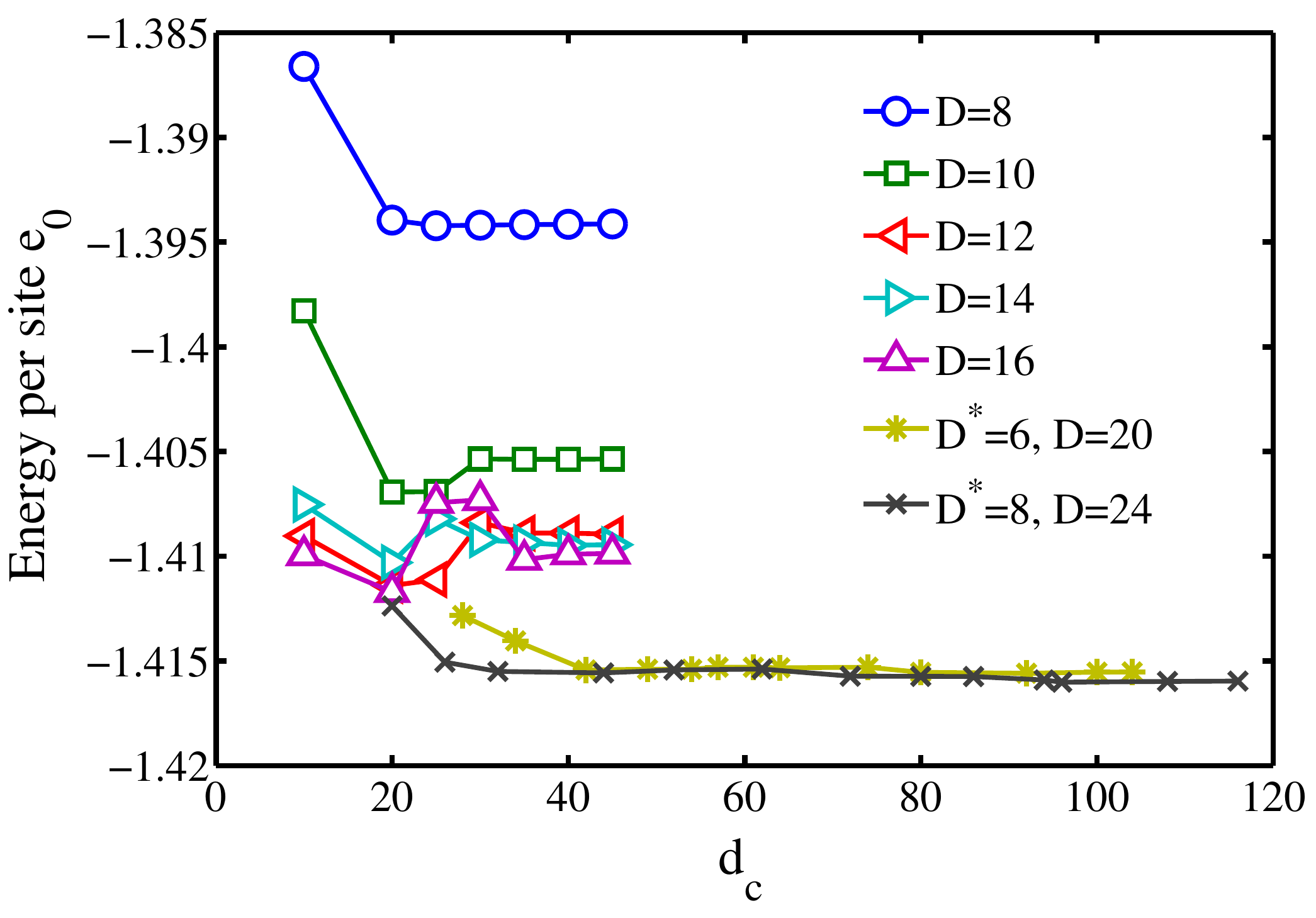}
\caption{(Color online) The convergence of $e_0$ with truncation parameters $d_c$ of iPEPS contractions. The two lines of $D^*=6$ and $D^*=8$ both have the same $y$-axis offset of $-0.005$, in order to not clutter the plot.}
\label{figS4}
\end{figure}

{\section{Convergence of energy per site $e_0$ versus truncation parameter $d_c$}}
In our iPEPS contractions (both with and without SU(2) symmetry implemented), the convergence of calculated energy per site $e_0$ versus the truncation parameter, i.e., the bond dimension $d_c$ of boundary MPS, are always checked. In Fig. \ref{figS4}, we show that the convergence behavior of $e_0$ versus $d_c$, for various PEPS bond dimensions $D$. From Fig. \ref{figS4}, one can observe that, when  $D$ becomes larger (say, $D>10$), larger $d_c$ are accordingly needed for the results to converge. Thanks to the implementation of SU(2) symmetry, we are able to retain $d_c = 110\sim120$ states ($d_c^*=40\sim45$ multiplets) on the geometric bond of boundary MPS, which is necessary for the accurate contraction of $D\geq20$ PEPS.

\begin{thebibliography}{99}
\bibitem{fn-symm} Using the PEPS construction, geometric bonds are associated with auxiliary orthonormal state spaces that can be categorized according to the symmetries of the system. Through tensors, these state spaces are fused with others, as indicated by the arrows with all lines. Since the direction of arrows can be changed throughout the algorithm, there are no arrows shown with the lines in panel 3(a). Details will be published elsewhere.
\end{thebibliography}
\end{document}